\documentclass[preprint,12pt]{elsarticle}

\usepackage{lineno}
\usepackage{setspace}
\usepackage{amsmath}
\usepackage{graphicx}
\usepackage{subcaption}
\usepackage{float}
\usepackage{geometry}
\usepackage{textcomp}
\usepackage{microtype}
\usepackage[section]{placeins}
\usepackage[charter]{mathdesign}
\journal{Nuclear Instruments and Methods in Physics Research Section B}

\begin{document}

\begin{frontmatter}

\title{A comparison of simulation tools for Muon-Induced X-ray Emission (MIXE) in thin films: a study case with lithium batteries}

\author[psi]{Maxime Lamotte}
\author[psi]{Michael W. Heiss}
\author[psi]{Thomas Prokscha}
\author[psi]{Alex Amato}

\address[psi]{PSI Center for Neutron and Muon Sciences (CNM), 
Paul Scherrer Institute, 5232 Villigen PSI, Switzerland}

\begin{abstract}
We present a comparative study of three Monte Carlo simulation frameworks—SRIM, GEANT4, and PHITS—for modeling the transport, stopping, and atomic cascade of negative muons in micrometer-scale, multilayer systems relevant to Muon-Induced X-ray Emission (MIXE) experiments at the Paul Scherrer Institute (PSI). Using a lithium-ion battery as a benchmark target, simulated stopping depth profiles are compared with experimental data from the GIANT spectrometer. All three codes reproduce the overall muon depth distributions with good consistency, even across sharp density contrasts. SRIM provides reliable stopping depth estimates for compact geometries, whereas PHITS reproduces GEANT4 results with comparable accuracy and additionally generates muonic X-ray spectra. These spectra, however, exhibit a systematic energy offset in the K-line transitions of medium- and high-Z elements relative to theoretical and experimental values. Despite this bias, PHITS accurately captures relative intensities and spectral shapes, enabling element-specific line identification. The results demonstrate that SRIM and PHITS constitute practical tools for rapid estimation of muon stopping depth and stopping profiles, and that PHITS holds strong potential for predictive MIXE spectroscopy once its transition-energy bias is corrected. 
\end{abstract}


\end{frontmatter}


	\section{Introduction, MIXE at Paul Scherrer Institute}

\begin{figure}[H]
	\centering
	\includegraphics[width=0.7\linewidth]{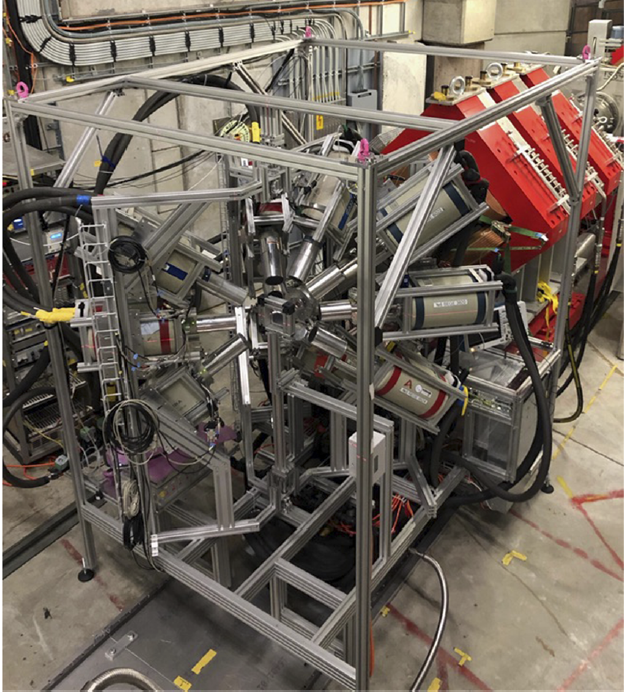}
	\caption{The GIANT setup for MIXE at the $\pi$E1 beamline at PSI. The muon beam is transported in the vacuum chamber on the top-right, and shaped by a triplet of quadrupole magnets (red) in order to focus the muons onto the sample, located in the center of the High-Purity-Germanium (HPGe) detector array. }
	\label{fig:giant}
\end{figure}

%
%
Muon-Induced X-ray Emission (MIXE) has long been employed to probe atomic-scale properties such as nuclear size, shape and charge distribution \cite{anderson1969,muX2021}. The field originated in the 1960s–1970s through precision muonic-atom spectroscopy, where measurements of transition energies revealed fine nuclear-structure effects beyond the reach of conventional X-ray methods. 
In the 1980s, pioneering work at the Swiss Institute for Nuclear Research (SIN)\textemdash now the Paul Scherrer Institute (PSI)\textemdash  extended MIXE to non-destructive elemental analysis of cultural and geological samples \cite{MIXE_PSI_1981,MIXE_PSI_1981_2}.  With PSI’s high-intensity continuous muon beams, generated at the High-Intensity-Proton-Accelerator-Facility HIPA \cite{HIPA_1}, the Swiss Muon Source (S\textmu S) now enables routine MIXE studies of bulk specimens using the GIANT HPGe array \cite{GIANT_2023} (Fig.~\ref{fig:giant}). Recent work conducted within the framework of the Swiss National Science Foundation (SNSF) interdisciplinary research program DEEP\textmu\ (SNSF Sinergia grant 193691) has enabled the application of MIXE across a broad range of fields—similarly to initiatives undertaken at other facilities—including depth-resolved, isotope-specific investigations of archaeological artifacts\cite{cataldo2022novel,biswas2023,hillier2025coin}, scientifically significant objects\cite{rossini2023meteorite,terada2014meteorite,ninomiya2019_lead_isotopes}, and energy-storage devices\cite{Izumi_2020,edouard}. These developments have further emphasized the need for robust simulation tools to support experimental design and data interpretation~\cite{cataldo2025MonteCarlo}.

In the following, $\mu$ denotes a negative muon ($\mu^-$) unless stated otherwise. MIXE relies on $\mu^-$ atomic capture and the ensuing muonic-atom cascade; positive muons ($\mu^+$) do not form bound muonic atoms and are therefore used primarily for $\mu$SR-type measurements rather than muonic X-ray emission.

Depth-resolved analysis is achieved by tuning the incident muon momentum, thereby shifting the stopping distribution within the sample without destructive sectioning. The achievable depth resolution ranges from a few micrometers to about 1~mm, depending on the relative momentum spread ($\Delta p/p$), the material density, and the stopping depth. The beam spot typically exhibits a waist diameter of about 2~cm, which varies with momentum, as lower-momentum muons are more susceptible to scattering in entrance detectors, insulation windows, and air.

For layered or encapsulated systems, MIXE is complementary to conventional ion-beam techniques such as Rutherford Backscattering Spectrometry (RBS) and Nuclear Reaction Analysis (NRA), as well as synchrotron-based X-ray methods. The high energy of muonic X-rays (typically 0.1 up to 10~MeV) reduces self-absorption compared to conventional XRF and enables analysis through thick overlayers or shells. Unlike Neutron Activation Analysis (NAA), the sample is not activated and can be returned shortly after measurement. A limitation is the cm-scale beam spot, implying limited lateral resolution compared with micro-focused ion or synchrotron probes; MIXE is therefore best viewed as a complementary tool for non-destructive, depth-resolved bulk analysis of complex objects.

\begin{figure}[H]
	\centering
	\includegraphics[width=0.8\linewidth]{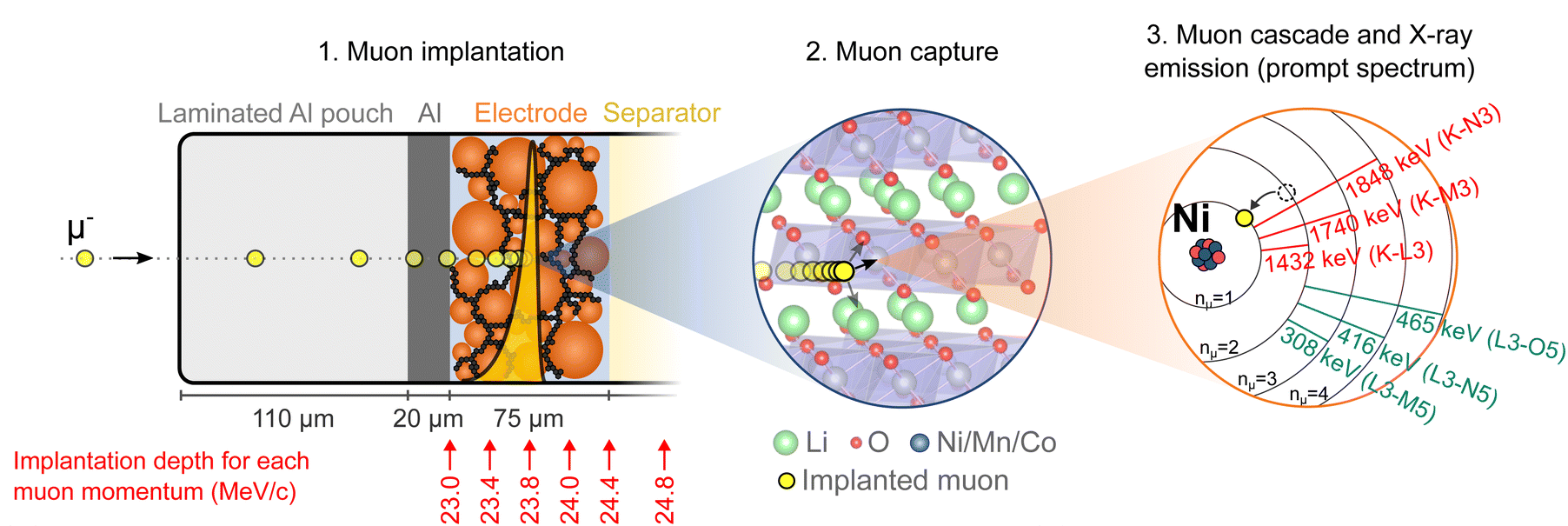}
	\caption{Steps to be modeled in PHITS: (1) muon transport from the beamline to the sample; (2) capture within the Coulomb field of a target atom; (3) muonic cascade with simultaneous emission of muonic X-rays. Image extracted from the work of Qu\'{e}rel et al. \cite{edouard}}
	\label{fig:schem_spectra_nickel}
\end{figure}

At the end of the stopping process in the sample, muons are captured in the Coulomb field of the positive charge of the target nucleus, creating a so-called muonic atom~\cite{anderson1969,muX2021,MIXE_PSI_1981}. This capture will leave the muon-atom system in an excited state with the principal quantum number $n_\mu \approx 14$~\cite{KNECHT_2020}, which subsequently decays to the ground state within a picosecond resulting in the emission of characteristic X-rays. This emission process can be compared to electronic X-ray fluorescence, but at two orders of magnitude higher energies due to the 207 times larger muon mass. 
The physics of muonic atoms and cascade processes is extensively described elsewhere \cite{hillier2022muon, amato2024Book}.
Fig.~\ref{fig:schem_spectra_nickel} depicts the 3 main processes of MIXE: muon stopping, muon capture, and muonic cascade.

Computer codes such as Akylas~\cite{Akylas_code} or MuDirac~\cite{MuDirac_2021} can produce an extended database of typical X-ray photons originating muonic cascade, taking into account both isotopic and orbital differences. PHITS~\cite{mandatoryPHITS,phits_333,phits_benchmark} and GEANT4~\cite{mandatoryG4, cataldo2023G4MIXE} can compute these typical transitions and output a tentative muonic X-ray photon spectrum.



In the MIXE technique at PSI, muons delivered by the $\pi$E1 beamline are implanted into the sample with adjustable momenta between 15 and 60~MeV/c, corresponding to a maximum penetration of about $\mathrm{3~g/cm^2}$. An upgrade of the $\mu$E1 beamline is planned to extend this range up to 125~MeV/c, enabling substantially deeper stopping depths. This beamline transports muons from in-flight pion decay produced in the thick graphite Target~E station at HIPA \cite{HIPA_1} through a long superconducting magnet. In the upgraded configuration, momenta between 60 and 125~MeV/c will be selectable, corresponding, for example, to penetration depths from the millimeter scale up to approximately 10~cm in aluminium.

The sensitivity of MIXE to light isotopes (Na, Li) was first demonstrated in Japan~\cite{Izumi_2020,Izumi_2018}, and this result enabled a team from the Swiss Federal Laboratories for Material Sciences and Technology (Empa) to begin applying MIXE for in situ following of electrochemical processes inside battery cells~\cite{edouard}. This experimental setup and data and will be implemented in the simulation frameworks and compared to numerical models described below.

%
%

%
%
%
%

The objective of this work is therefore to identify simulation tools that are both reliable and practical for predicting muon stopping profiles and muonic X-ray spectra in layered materials, with particular emphasis on usability for experiment preparation. 
The benchmarking strategy adopted here (transport/stopping validation and spectral-model validation) is not specific to lithium-ion batteries and can be transferred to other heterogeneous layered or encapsulated systems such as cultural-heritage coatings/corrosion layers, geological/extraterrestrial specimens, and solid-state battery architectures.
Using a multilayer lithium-ion battery as a representative heterogeneous target with demanding simulation requirements, we benchmark SRIM, GEANT4, and PHITS against each other and against experimental measurements. The study aims to determine the level of accuracy that can be expected from each code and to provide guidance for future users selecting simulation tools for MIXE-based investigations.
We emphasize that the present benchmark focuses on transport/stopping accuracy and on the internal consistency of cascade-generated spectra; it does not attempt a full detector-response model or an absolute-efficiency validated prediction of experimental count rates.
%
\section{Materials and methods}

\subsection{Experimental geometry: the GIANT setup}

The GIANT setup (Fig.~\ref{fig:giant}) consists of an aluminium frame hosting up to 30 high-purity germanium (HPGe) detectors, continuously cooled by a liquid nitrogen circulation system. Samples are analyzed in air, positioned 10–20~cm from the beamport window depending on detector placements. Muons from the High-Intensity Proton Accelerator (HIPA)~\cite{HIPA_1} are transported by the $\pi$E1 beamline and focused by a quadrupole triplet onto a plastic scintillator tagger (BC-400, 21×22~mm$^2$, 200~\textmu m thick). The light output is read by Advansid NUV-series SiPMs, providing $\sim$0.4~ns timing resolution. After tagging, muons cross a 10~\textmu m titanium window into air before reaching the sample. Multiple scattering in these entrance materials slightly broadens the beam and increases angular divergence at the target.

\subsection{Simulation tools and configuration}

Muon slowing-down and capture were modeled using SRIM~\cite{mandatorySRIM}, PHITS~\cite{mandatoryPHITS,phits_333,phits_benchmark}, and GEANT4~\cite{mandatoryG4,cataldo2023G4MIXE}.  
All three codes simulate muon deceleration via electronic and nuclear scattering to estimate stopping profiles within the layered lithium-ion battery cell described in Ref.~\cite{edouard}.

\noindent\textbf{SRIM.}\\[0.3em]
Since SRIM does not natively support muons, pseudo-muons were modeled as protons with rescaled mass $m_\mu = 105.7/931.5 \simeq 0.113$~amu.  
Because SRIM’s stopping powers scale with $Z^2$ and rely on Coulomb scattering, this approximation reproduces muon energy loss reasonably well while avoiding charge-exchange artifacts.  
Despite its limitations, SRIM (and TRIM.SP~\cite{Biersack1984}, an ion transport and sputtering code) has been successfully applied to positive-muon transport in thin films~\cite{TRIM_Elvezio}, and here serves as a lightweight benchmark for MIXE energies. Importantly, SRIM offers no possibility to simulate realistic beam optics: only monoenergetic, pencil-like beams incident on semi-infinite targets can be used.  

\begin{figure}[htb] 
	\begin{subfigure}[b]{0.565\linewidth}
		\centering
		\includegraphics[width=0.95\linewidth]{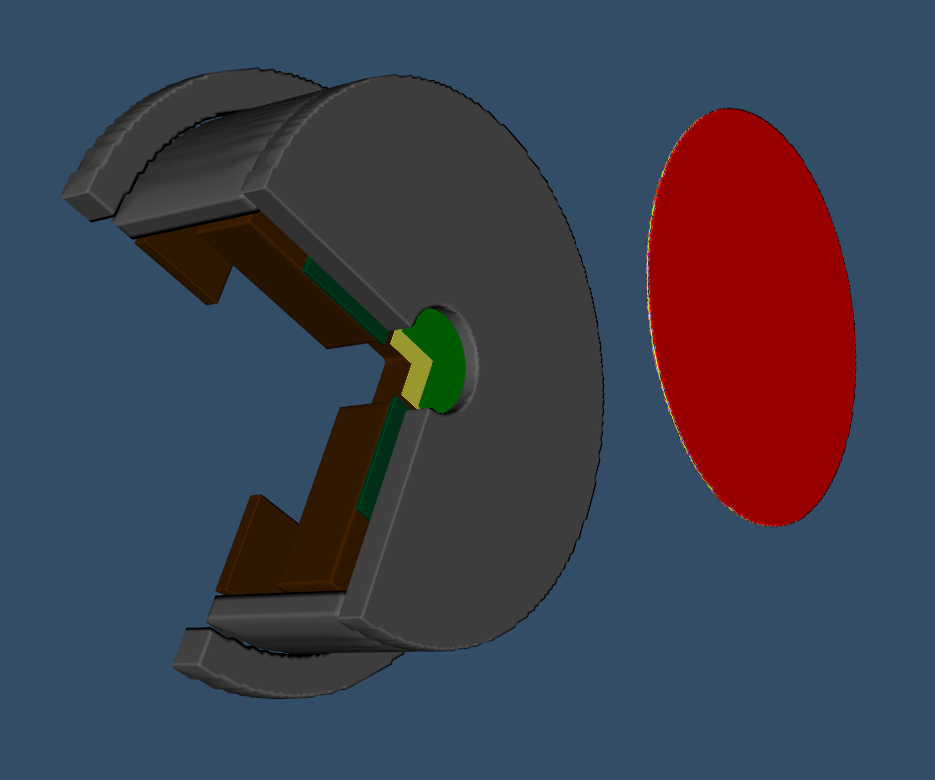}
	\end{subfigure}%
	\begin{subfigure}[b]{0.435\linewidth}
		\centering
		\includegraphics[width=0.95\linewidth]{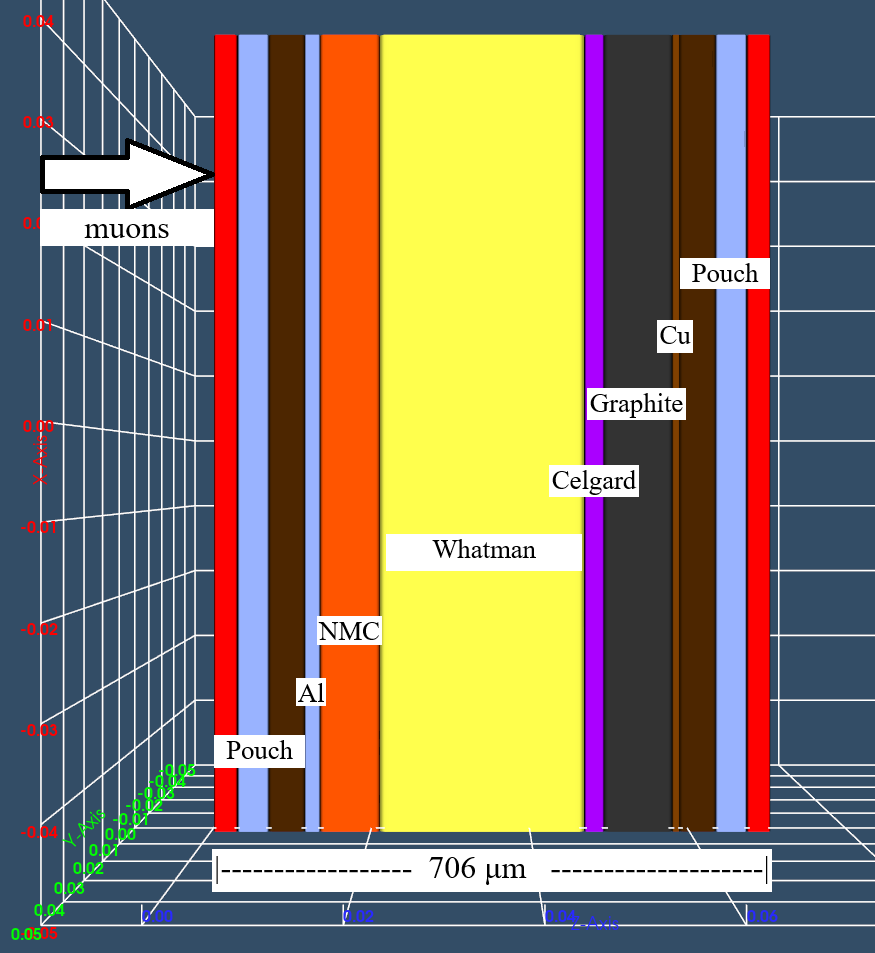}
	\end{subfigure}
	\caption{Left: PHITS geometry showing the muon beam (from left) through the tagger (yellow) and titanium window (green) before reaching the sample (red) 14~cm downstream. Right: Zoom on the multilayer battery stack. Material parameters are listed in Tab.~\ref{tbl:composition}.}
	\label{fig:3d}
\end{figure}

\noindent\textbf{PHITS and GEANT4.}\\[0.3em]
Recent improvements in the PHITS muon-interaction module~\cite{PHITS_muon} motivated its benchmarking against GEANT4, the reference toolkit for particle transport.  
Muon capture cross sections were taken from Suzuki \textit{et al.}~\cite{CaptureXS_Suzuki_1987}, and the ensuing muonic X-ray cascade was simulated using the Akylas–Vogel routine~\cite{Akylas_code}.  
Each code modeled the same simplified geometry (Fig.~\ref{fig:3d}), including the tagger, vacuum window, 10~cm air gap, and eight battery layers with densities and thicknesses listed in Tab.~\ref{tbl:composition}.  
Beam momentum spreads of $\pm$0.5\% around nominal momenta were implemented to reflect the $\pi$E1 beam characteristics.  A nominally pencil-like muon beam was generated in both codes; however, the multiple scattering induced by the muon tagger results in a significantly broadened beam spot after 10~cm of air gap, effectively defining the incident profile on the sample. 

\noindent All simulations were conducted 
using SRIM-2008/TRIM~2013, GEANT4~v11.2.1, and PHITS~v3.34.

\begin{table}[htb]
\centering
\caption{Thickness and density of the battery layers. The pouch is a polyamide–Al–polyethylene composite. Here, ``Whatman'' denotes a glass-fiber separator, whereas ``Celgard'' denotes a thin microporous polymer separator film.}
\vspace{2mm}
\begin{tabular}{lcccccccc}
\hline
Layer & Pouch & Al & NMC & Whatman & Celgard & Graphite & Cu & Pouch \\
\hline
Density (g/cm$^3$) & 1.54 & 2.70 & 3.38 & 1.30 & 1.05 & 1.87 & 8.92 & 1.60 \\
Thickness (\textmu m) & 115 & 20 & 75 & 260 & 25 & 86 & 10 & 115 \\
Depth (\textmu m) & 115 & 135 & 210 & 470 & 495 & 581 & 591 & 706 \\
\hline
\end{tabular}
\label{tbl:composition}
\end{table}

\subsection{Simulation scope and precision}

PHITS offers multiple physics options for transport accuracy; the key cards used are listed in Appendix~A.  
A sensitivity study showed that the \texttt{nspred} (Coulomb diffusion) flag alters stopping depth by less than 1\%, indicating minimal angular-straggling influence under the present conditions. The GEANT4 application uses physics library QGSP\_BIC\_HP\_EMZ and G4 material database.

Statistical uncertainties on simulated observables were evaluated from counting statistics within each histogram bin assuming Poisson behavior and propagated to derived quantities (mean stopping depth, peak intensities, and fitted Gaussian areas). In the present simulations ($\sim10^{6}$ primaries), the standard error of the mean (SEM) on the mean stopping depth (MSD) is sub-\textmu m and does not limit the GEANT4/PHITS comparison. Additional systematic contributions may arise from histogram binning choices, detector response assumptions used for peak integration, from imperfect knowledge of the experimental geometry (layer thicknesses, material densities, air gaps and alignment), and detector efficiency. Manufacturer tolerances for the benchmark cell were not available and these geometry-related uncertainties are therefore not quantified here; they should be regarded as a potential dominant contribution when comparing to experimental data.
Output stopping depth profiles were averaged over the final 5\% of trajectories to derive the mean range and standard deviation.  
The results are compared quantitatively in the next section.

%
%
\section{Results and discussion}
\subsection{Muon stopping depths}
\subsubsection{Momentum scan and stopping-depth comparison}

A total of 56 muon momenta were investigated to determine the corresponding penetration and stopping depths within the modeled battery cell. The lower momentum bound was defined by the minimum value at which muons reach the target, as a significant fraction is otherwise stopped in the titanium window, while the upper bound corresponds to momenta for which muons traverse the entire stack without stopping.

For both GEANT4 and PHITS, momenta between 20~MeV/c and 30~MeV/c were simulated with up to $2\times10^{6}$ primary particles, following preliminary runs of $10^{4}$ events to establish the limits of the momentum scan. SRIM results were obtained using $10^{4}$ pseudo-muons. The MSD calculated with each code is summarized in Tab.~\ref{tbl:depths}.

%
\subsubsection{Uncertainty estimates}

To quantify the statistical reliability of the simulated stopping depths, the MSD was computed from the stopping-position distribution and its spread (range straggling) defined as:
\begin{align}
    \sigma = \sqrt{\frac{\Sigma_i w_i (x_i - MSD)^2}        {\Sigma_i w_i}}
\end{align}
where $x_i$ denotes the depth coordinate and $w_i$ the number of stopped muons in bin index $i$. The statistical uncertainty on the MSD, the SEM, was estimated as
\begin{align}
    SEM = \frac{\sigma}      {\sqrt{N_{eff}}}
\end{align}
with $N_{\mathrm{eff}}=\sum_i w_i$ the effective number of stopped muons. As mentioned, the resulting statistical uncertainties are below 0.1~\textmu m and therefore negligible compared to the intrinsic stopping depth spread $\sigma$, which is typically several to a few tens of micrometers, depending on material and momentum, as compiled in Tab.\ref{tbl:ranges_air}. When comparing to experimental spectra, additional systematic contributions (geometry, detector efficiency, background subtraction, and low-energy attenuation) can dominate and are discussed separately in Sec.~\ref{sec:Xray_from_muon}.

%
\begin{table}[htb]
\small
\centering
\caption{Mean stopping depths (in \textmu m) in the battery, computed by GEANT4 (baseline), PHITS, and SRIM. The SEM on the Monte Carlo mean depth is $<0.1$~\textmu m for GEANT4 and PHITS at all momenta for the present statistics.}
\vspace{3mm}
\begin{tabular}{cccccccccccc}
P~(MeV/c)& 20 & 21 & 22 & 23 & 24 & 25 & 26 & 27 & 28 & 29 & 30 \\
\hline
\hline
GEANT4 & 12.9 & 50.9 & 102.1 & 145.7 & 177.7 & 220.4 & 308.8 & 413.2 & 518.3 & 599.7 & 672.5 \\
PHITS  & 13.0 & 49.0 & 96.8  & 146.0 & 176.5 & 216.9 & 309.5 & 405.9 & 510.4 & 595.7 & 675.1 \\
\textbf{$\Delta$ (\%)} & \textbf{0.8} & \textbf{-3.7} & \textbf{-5.2} & \textbf{0.2} & \textbf{-0.7} & \textbf{-1.6} & \textbf{0.2} & \textbf{-1.8} & \textbf{-1.5} &\textbf{ -0.7} & \textbf{0.4 }\\
SRIM   & air  & air  & air   & 114   & 155   & 192   & 203   & 339   & 368   & 563   & 644   \\
\hline
\end{tabular}
\label{tbl:depths}
\end{table}
While GEANT4 and PHITS results are nearly identical, SRIM exhibits significant discrepancies, especially when large air volumes precede the thin-film stack. To isolate the origin of this discrepancy, the range of pseudo-muons in each battery layer was recalculated with SRIM and PHITS (Tab.~\ref{tbl:ranges}).
%
\subsubsection{SRIM limitations in low-density media}

Previous studies\cite{cataldo2023Air} have reported that SRIM range predictions can be improved when extended low-density regions are replaced by compressed-density equivalents, effectively reducing numerical precision limitations associated with large path lengths. While this approach can reproduce realistic stopping positions, it alters the physical geometry and therefore was not adopted here, since the goal of this work is a direct comparison of transport codes under identical and physically faithful conditions. We note that such density scaling may be useful for rapid range estimation but does not remove intrinsic precision limitations when long trajectories must be explicitly tracked.
%
\begin{table}[htb]
\small
\centering
\caption{Mean stopping depth in \textmu m of muons in individual battery layers at various momenta $P$, computed by PHITS (baseline) and SRIM. The SEM on the Monte Carlo mean stopping depth is $<0.1$~\textmu m for PHITS at all momenta for the present statistics.\label{tbl:ranges}}
\begin{tabular}{c c c |c c|c c|c c|c c|c c|c c}
Layer & \multicolumn{2}{c|}{Pouch} & \multicolumn{2}{c|}{Al} &  \multicolumn{2}{c|}{NMC} & \multicolumn{2}{c|}{Whatman} & \multicolumn{2}{c|}{Celgard} & \multicolumn{2}{c|}{Graphite}& \multicolumn{2}{c}{Cu}   \\
P (MeV/c)& 20 & 30 & 20 & 30 & 20 & 30 & 20 & 30 & 20 & 30 & 20 & 30 & 20 & 30 \\
\hline
\hline
PHITS & 203 & 876 & 174 & 720 & 117 & 496 & 279 & 1193 & 331 & 1410 & 196 & 840 & 65 & 264 \\
SRIM  & 205 & 873 & 169 & 700 & 115 & 485 & 274 & 1180 & 330 & 1410 & 196 & 831 & 63 & 252 \\
\textbf{$\Delta$ (\%)} & \textbf{1.0} & \textbf{-0.3} & \textbf{-2.9} & \textbf{-2.8} & \textbf{-1.7} & \textbf{-2.2} & \textbf{-1.8} & \textbf{-1.1} & \textbf{-0.3} & \textbf{0.0} & \textbf{0.0} & \textbf{-1.1} & \textbf{-3.1} & \textbf{-4.5} \\
\end{tabular}
\end{table}
For isolated layers, SRIM and PHITS agree within 5\%. Additional tests on upstream materials (scintillator, titanium window, and various gases) confirmed this consistency (Tab.~\ref{tbl:ranges_air}). However, we note a larger SRIM-computed spread in titanium and scintillator.
An important difference between PHITS and SRIM algorithms arises from the treatment of negative muon nuclear capture. SRIM/TRIM does not include muon capture physics and therefore describes purely electromagnetic slowing down and stopping. 

\subsubsection{Effect of muon capture in upstream materials}
In PHITS, when negative muon capture is enabled, a fraction of the muons are absorbed by the nucleus shortly after atomic capture, modifying the spatial distribution of disappearance points. As muonic capture probability increases with $Z$, we additionally repeated the PHITS runs with muon capture \textit{imucap} disabled in the last column of Tab.~\ref{tbl:ranges_air} to isolate purely electromagnetic stopping and assess its impact on the disappearance-depth distribution.
\begin{table}[htb]
\small
\centering
{\caption{Mean stopping depth and stopping-depth spread ($\sigma$, i.e.\ range straggling) for selected upstream materials as computed by SRIM and PHITS, with and without the negative-muon capture flag \textit{imucap}. The SEM on the Monte Carlo mean depth is $<0.1$~\textmu m for PHITS at all momenta for the present statistics.}\label{tbl:ranges_air}}
\begin{tabular}{c c c c c c c c}
Layer & P (MeV/c) & PHITS (w. mu) & $\sigma$ & SRIM & $\sigma$  & PHITS (w/o mu) & $\sigma$\\
\hline
He (cm)      & 20  & 189 & 7.33         & 192 & 7.37    & 189  & 7.27   \\
He (cm)      & 30  & 829 & 2.80         & 825 & 3.20   & 830 & 2.87    \\
Air (cm)    & 20  & 32.1 & 1.18         & 34.0 & 1.91   & 32.3 & 1.55   \\
Air (cm)     & 30  & 137 & 5.37         & 138  & 6.43   & 135 & 5.05   \\
CO$_2$ (cm)  & 20  & 21.0 & 0.73        & 21.4 & 1.01   & 20.6 & 0.89   \\
CO$_2$ (cm) & 30  & 88.0 & 6.46         & 90.1 & 4.17   & 87.2 & 5.75   \\
BC400 (\textmu m)  & 20  & 325 & 1.74   & 330  & 16.0   & 325  & 1.78   \\
BC400 (mm)         & 30  & 1.40& 0.06   & 1.42 & 0.06   & 1.40 & 0.06   \\
Ti (\textmu m)     & 20  & 117 & 2.80   & 114  & 8.68   & 117  & 2.82   \\
Ti (\textmu m)     & 30  & 478 & 9.43   & 465  & 33.0   & 476  & 9.43   \\
\end{tabular}
\end{table}

\begin{figure}[htb]
\centering
\includegraphics[width=0.99\linewidth]{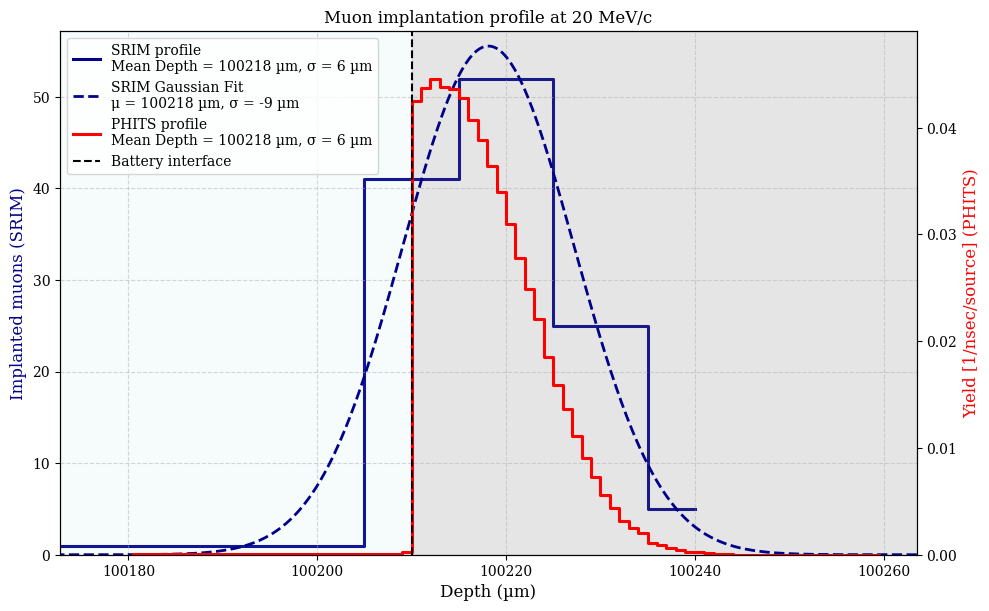}
\caption{20~MeV/c muon stopping depth profile as computed by SRIM (blue) for a 10~cm air cell (light blue) before the battery. The Gaussian fit (dashed line) yields 100218~\textmu m ($R^2$ = 0.97), while SRIM’s internal mean is 97570~\textmu m. PHITS (red) highlights the density transition at the first PE layer (light grey).}
\label{fig:gaussian_SRIM}
\end{figure}

%
The discrepancies in Tab.~\ref{tbl:depths} stem from SRIM’s averaging procedure. When a large low-density region precedes the target (e.g., air), the mean depth is dominated by those trajectories. Fitting the stopping depth tail with a Gaussian yields more realistic ranges consistent with PHITS, as shown in Fig.~\ref{fig:gaussian_SRIM}. The limited floating-point precision (5 digits, fixed \textmu m unit) also restricts SRIM’s spatial resolution to $\sim$10~\textmu m over a 10~cm setup.

\begin{figure}[H]
\centering
\includegraphics[width=0.95\linewidth]{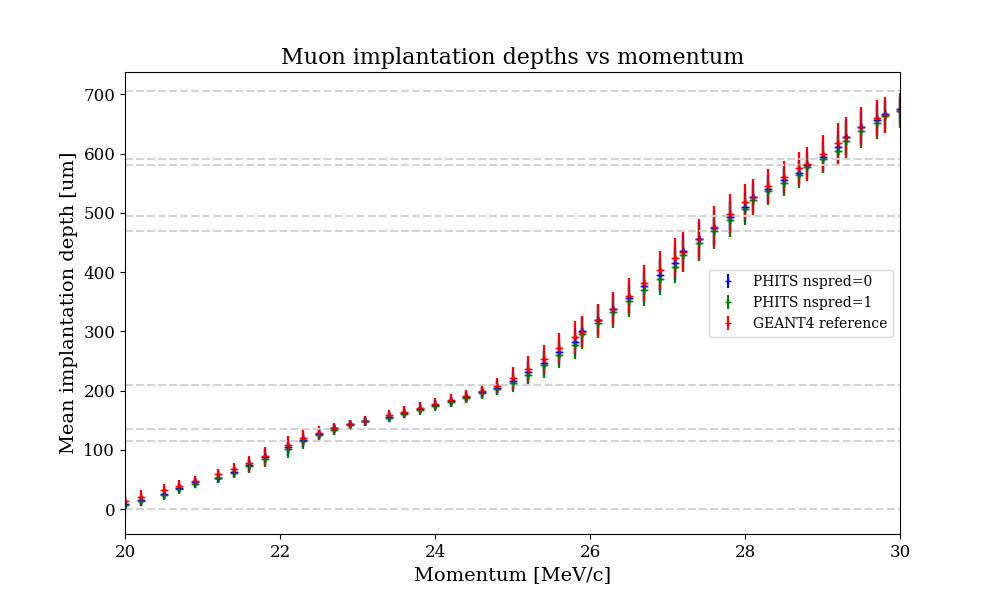}
\caption{Simulated mean stopping depths of muons in the battery cell. Dashed lines indicate layer interfaces. Vertical markers translate the standard deviations of the mean stopping depth.}
\label{fig:mean}
\end{figure}

Despite these limitations, SRIM remains useful for preliminary range estimates. Even without muon capture physics implemented, SRIM is capable of reproducing ranges of negative muons with an accuracy similar to PHITS with its extended muon physics routines, as seen in Tab.\ref{tbl:ranges_air}. For subsequent analyses, only GEANT4 and PHITS were retained. The mean stopping depths and distribution spread from both codes are shown in Fig.~\ref{fig:mean}. 
\subsubsection{Sensitivity to transport settings}
\begin{figure}[H]
\centering
\includegraphics[width=0.95\linewidth]{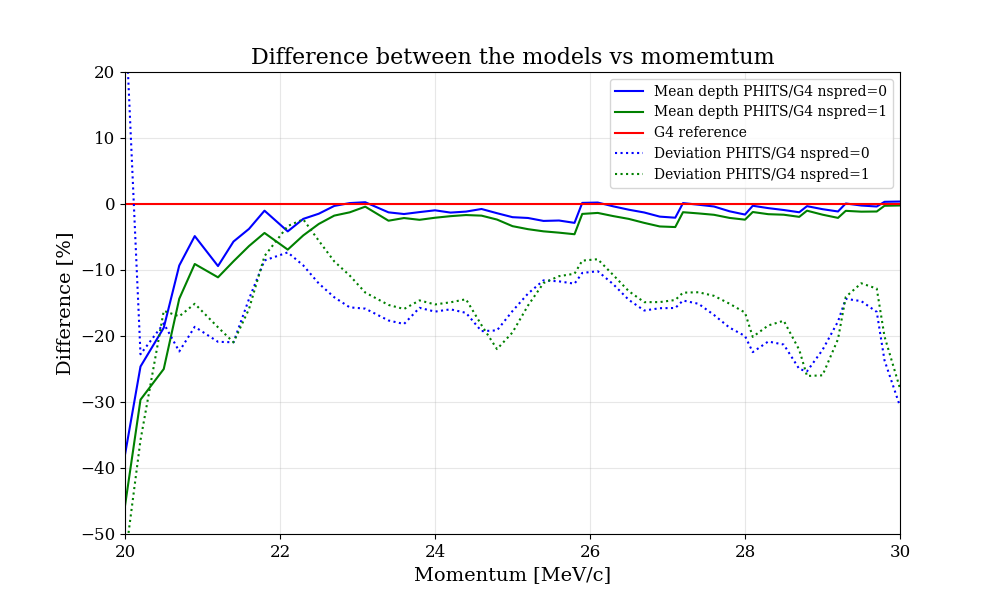}
\caption{Solid lines: relative differences in mean muon stopping depth between PHITS and GEANT4 (reference). Larger discrepancies occur at low momentum due to scattering in air. Dashed lines: relative differences in standard deviation of the mean stopping depth as function of momentum, between PHITS and GEANT4 (used as reference model).}
\label{fig:difference}
\end{figure}

Agreement at the percent level is obtained irrespective of the PHITS \textit{nspred} flag, indicating limited sensitivity to the corresponding angular straggling settings for this geometry. Momentum values between 23–28~MeV/c were thus selected for optimal probing between the Al and Cu electrodes. The residuals between codes (Fig.~\ref{fig:difference}) show discrepancies only at low momentum, dominated by scattering in upstream layers.

\begin{figure}[ht] 
	\begin{subfigure}[b]{0.33\linewidth}
		\centering
		\includegraphics[width=0.99\linewidth]{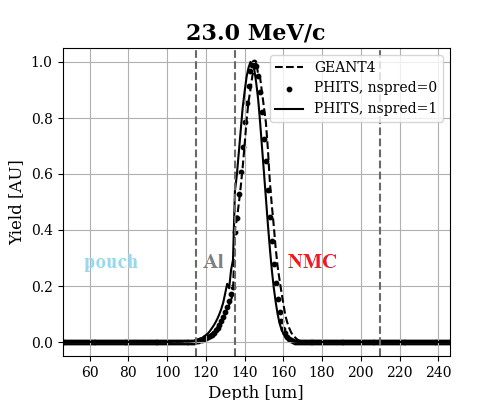} 
		\vspace{4ex}
	\end{subfigure}
	\begin{subfigure}[b]{0.33\linewidth}
		\centering
		\includegraphics[width=0.99\linewidth]{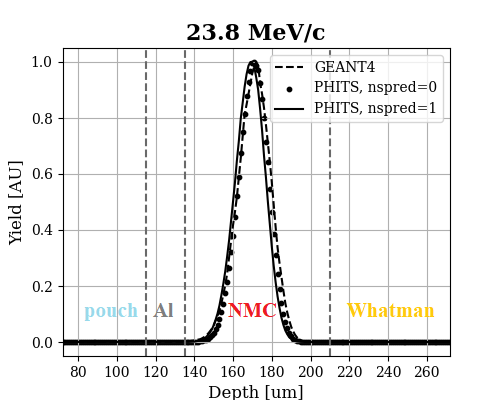} 
		\vspace{4ex}
	\end{subfigure} 
	\begin{subfigure}[b]{0.33\linewidth}
		\centering
		\includegraphics[width=0.99\linewidth]{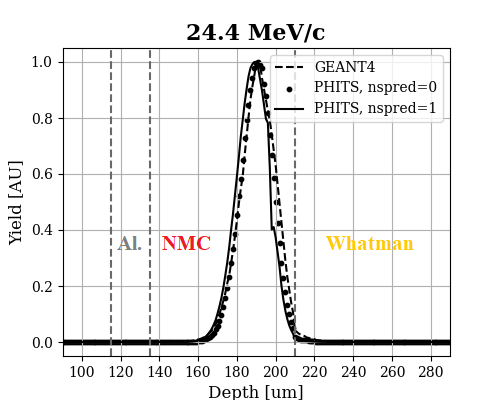} 
		\vspace{4ex}
	\end{subfigure} 
	\begin{subfigure}[b]{0.33\linewidth}
	\centering
	\includegraphics[width=0.99\linewidth]{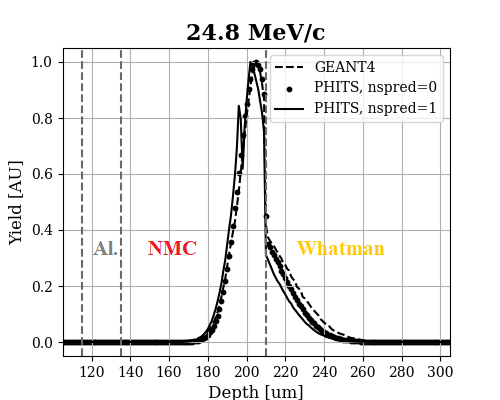} 
	\vspace{4ex}
\end{subfigure}
\begin{subfigure}[b]{0.33\linewidth}
	\centering
	\includegraphics[width=0.99\linewidth]{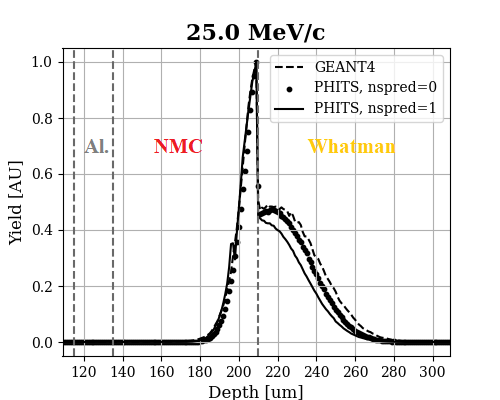} 
	\vspace{4ex}
\end{subfigure} 
\begin{subfigure}[b]{0.33\linewidth}
	\centering
	\includegraphics[width=0.99\linewidth]{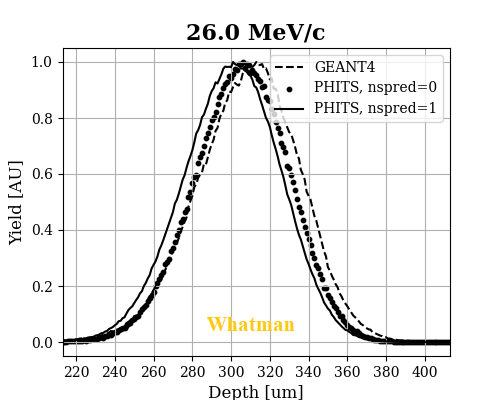} 
	\vspace{4ex}
\end{subfigure} 
	\begin{subfigure}[b]{0.33\linewidth}
	\centering
	\includegraphics[width=0.99\linewidth]{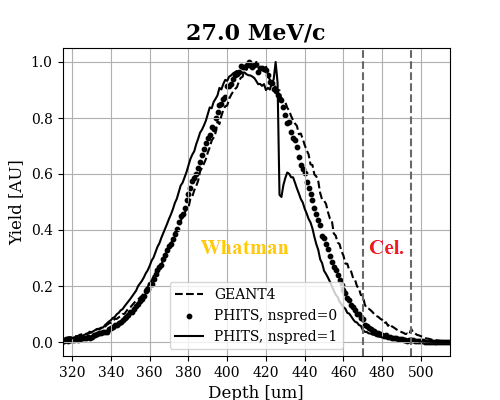} 
	\vspace{4ex}
\end{subfigure}
\begin{subfigure}[b]{0.33\linewidth}
	\centering
	\includegraphics[width=0.99\linewidth]{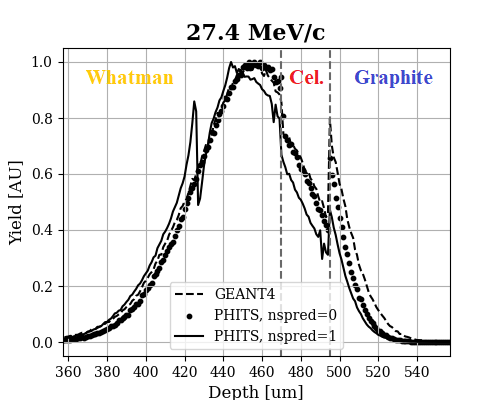} 
	\vspace{4ex}
\end{subfigure} 
\begin{subfigure}[b]{0.33\linewidth}
	\centering
	\includegraphics[width=0.99\linewidth]{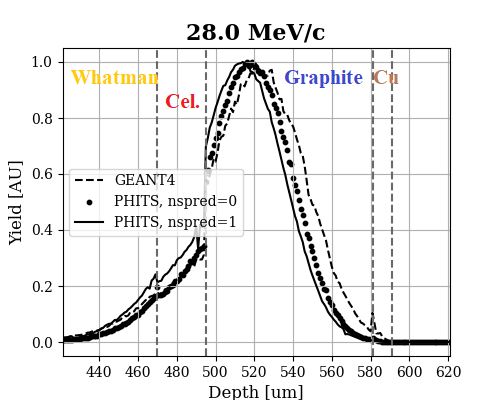} 
	\vspace{4ex}
\end{subfigure} 	
\caption{stopping depth profiles of muon as a function of beam momentum. Results are presented with the GEANT4 results as a reference (dashed line) and for PHITS having the physics card nspred = 0 (dots) and nspred = 1 (solid line). Graphs are labeled with corresponding muon beam momenta. Muons come from the left, as represented in Fig.~\ref{fig:3d}}
\label{fig:profiles_all}
\end{figure}

The stopping depth profiles for several momenta (Fig.~\ref{fig:profiles_all}) illustrate these effects. Anomalous discontinuities (e.g., around 425~\textmu m in the Whatman layer at 27.4~MeV/c) appear only with \textit{nspred}=1 and result from numerical inconsistencies in the slowing-down calculations of PHITS.

\subsection{X-ray emission from muon cascades} \label{sec:Xray_from_muon}

\subsubsection{Spectral overview and transition energy discrepancies}

\begin{figure}
\centering
\includegraphics[width=1.0\linewidth]{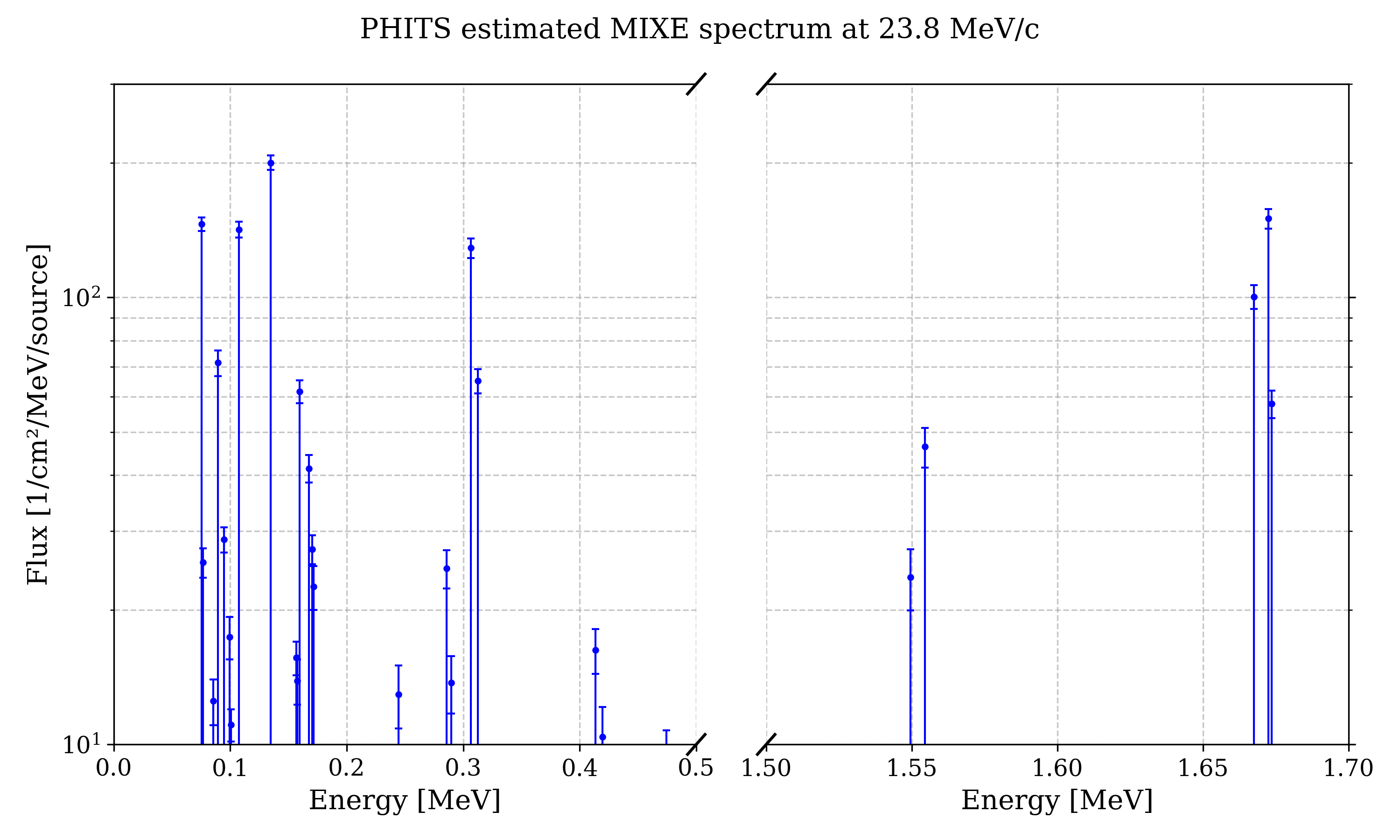}
\caption{Raw MIXE emission spectrum from PHITS at 23.8~MeV/c. Low-energy peaks correspond to light elements K-lines MIXE signal and heavier elements (Mn, Ni, Co) L-lines. High-energy peaks corresponding to K-lines of the latter show shifted positions compared to reference data. The error bars represent the one-standard-deviation (1~$\sigma$) statistical uncertainties provided by PHITS for each energy bin. These correspond to the relative Monte Carlo tally error (r.err) multiplied by the bin flux value (1~keV wide), reflecting counting statistics from the finite number of simulated particle histories.}
\label{fig:mixeraw}
\end{figure}

PHITS-computed photon spectra exhibit typical MIXE lines (Fig.~\ref{fig:mixeraw}), with low-energy $K_{\alpha}$ transitions (C, Si, Al) well reproduced, while heavier elements (Ni, Mn, Co) show energy-biased peaks. This issue is systematic across datasets.
%

\begin{figure}
\centering
\includegraphics[width=1\linewidth]{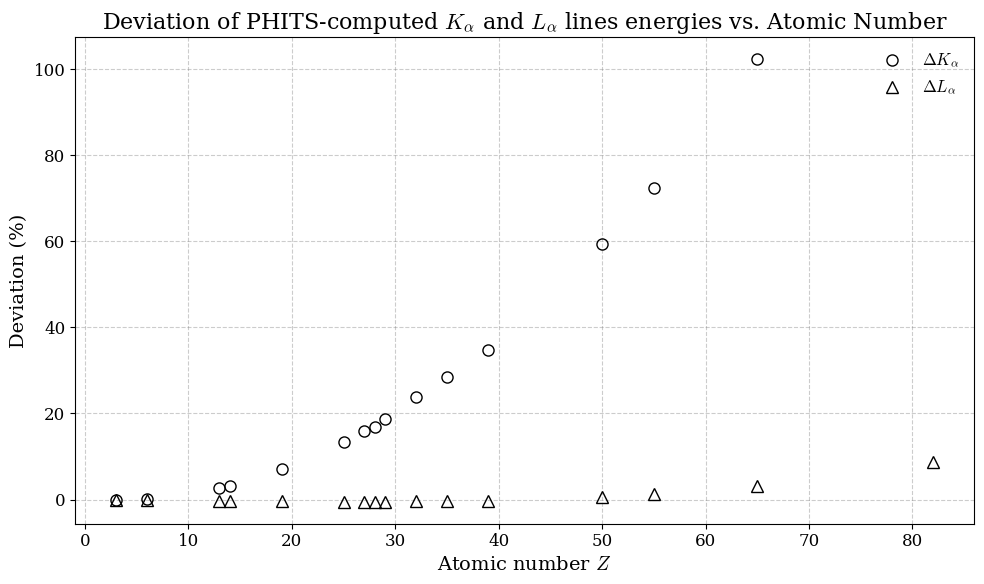}
\caption{Relative energy difference between PHITS and MuDirac for K and L transitions. K-lines are more affected than L-lines, with the discrepancy increasing with $Z$.}
\label{fig:shift}
\end{figure}

Comparison between PHITS (Akylas’ code) and MuDirac reveals an increasing energy discrepancy for high-$Z$ elements (Fig.~\ref{fig:shift}), reaching several hundred keV for Ni and Cu.
In the present PHITS version, the dominant contribution to this bias was traced to an energy scaling issue in the \textit{aama.f} cascade routine; more generally, low-$n$ (K-series) muonic levels are increasingly sensitive to relativistic and finite-nuclear-size corrections with increasing $Z$, which explains why such discrepancies tend to grow for heavier elements and affect K lines more strongly than L lines.
The relative difference is shown in Fig.~\ref{fig:shift}, with detailed values in Tab.~\ref{tbl:appendix_shift}. To pinpoint which peak is associated to each isotope, we ran single element test run with 16 different targets. A dump file, called by modifying PHITS executable, would output the detailed cascade generated by \textit{aamaa.f}, including transition levels, energies and intensities. 
To enable diagnostic dumping of the \textit{aama.f} cascade routine, hydrogen-bearing materials were temporarily replaced by helium in a dedicated technical test configuration to avoid a known instability of the PHITS atomic database interface. This substitution was used only for debugging the cascade bookkeeping and does not affect the transport/comparison results presented above.

\begin{figure}[ht]
\centering
\includegraphics[width=1\linewidth]{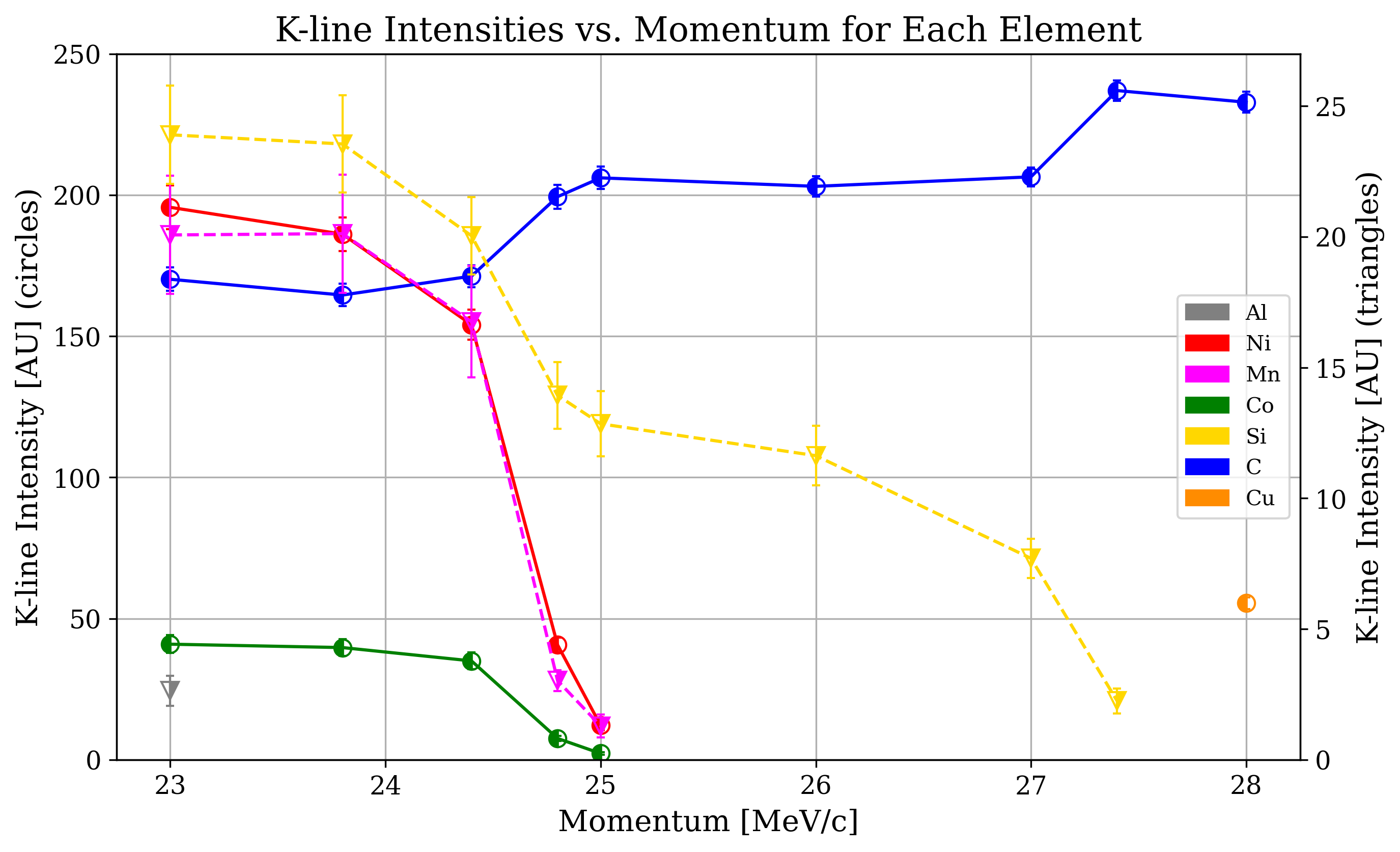}
\caption{PHITS calculated K-line intensities (arbitrary units) versus beam momentum. Low-intensity lines are magnified by a factor of ten (dashed). NMC materials can be identified at low momenta, as they are present in the first layers of the battery element. Cu peaks appear only at the highest momenta as this element is only present on the electrode placed downstream of muon path. A resolution of ±1~keV was applied for peak identification. Error bars denote $1\sigma$ statistical uncertainties of the PHITS tallies  propagated to the integrated peak intensities over the stated energy window.}
\label{fig:K-lines}
\end{figure}

After identifying these systematic offsets, the relevant peaks were isolated in the composite spectrum, shown in Appendix. The evolution of MIXE peaks with momentum reflects how the muons progressively penetrate deeper into the battery layers.
At low momentum, only Al lines are visible, followed by Ni, Mn, Co, and Si as muons penetrate deeper. Cu K-lines emerge only at 28~MeV/c because the muon stopping distribution reaches the Cu current collector located at the back of the multilayer stack; at lower momenta the stopping probability at that depth is negligible (see muon stopping depth profiles in Fig.~\ref{fig:profiles_all}). The evolution of K-line intensities with momentum is plotted in Fig.~\ref{fig:K-lines}. Error bars represent the 1~$\sigma$ statistical uncertainties of the PHITS Monte Carlo tallies. For each K-line, the uncertainty was obtained by propagating the per-bin relative error over a ±1~keV energy window. When multiple K-lines of the same element contribute at a given momentum, the plotted value corresponds to their mean intensity and the uncertainty is propagated assuming independent statistics. The quoted errors therefore reflect only counting (Monte Carlo) statistics.
While PHITS reproduces general spectral trends, absolute energies deviate from MuDirac, particularly for Ni and Cu isotopes (Tab.~\ref{tbl:iso_shift}). MuDirac solves the Dirac equation for a bound muon in a finite-size nucleus using realistic nuclear charge distributions (typically Fermi or homogeneously charged spheres), and includes vacuum-polarization and relativistic recoil corrections at leading order. The code provides transition energies with sub-keV accuracy for medium- and high-Z nuclei, and its predictions have been extensively benchmarked against precision muonic-atom X-ray measurements. PHITS underestimates isotopic energy shifts by nearly an order of magnitude compared to MuDirac and measurements.
\begin{table}[htb]
\centering
\caption{Energies in keV of some elements' muonic $\mathrm{K\alpha_1}$ and $\mathrm{K\alpha_2}$ transitions, as computed by MuDirac and PHITS.}
\vspace{3mm}
\begin{tabular}{cccc}
\hline
Isotope & Line & MuDirac & PHITS \\
\hline
$\mathrm{^{58}Ni }$& $\mathrm{K\alpha_1}$ & 1432.42 & 1672.99 \\
$\mathrm{^{58}Ni }$& $\mathrm{K\alpha_2}$ & 1426.90 & 1667.10 \\
$\mathrm{^{60}Ni }$& $\mathrm{K\alpha_1}$ & 1429.28 & 1673.10 \\
$\mathrm{^{60}Ni }$& $\mathrm{K\alpha_2}$ & 1423.75 & 1667.20 \\
$\mathrm{^{63}Cu }$& $\mathrm{K\alpha_1}$ & 1514.27 & 1796.66 \\
$\mathrm{^{63}Cu }$& $\mathrm{K\alpha_2}$ & 1507.95 & 1789.86 \\
$\mathrm{^{65}Cu }$& $\mathrm{K\alpha_1}$ & 1512.37 & 1796.76 \\
$\mathrm{^{65}Cu }$& $\mathrm{K\alpha_2}$ & 1506.05 & 1789.96 \\
\label{tbl:iso_shift}
\end{tabular}
\end{table}

\begin{figure}[htb]
    \begin{subfigure}[b]{1\linewidth}
        \centering
        \includegraphics[width=0.85\linewidth]{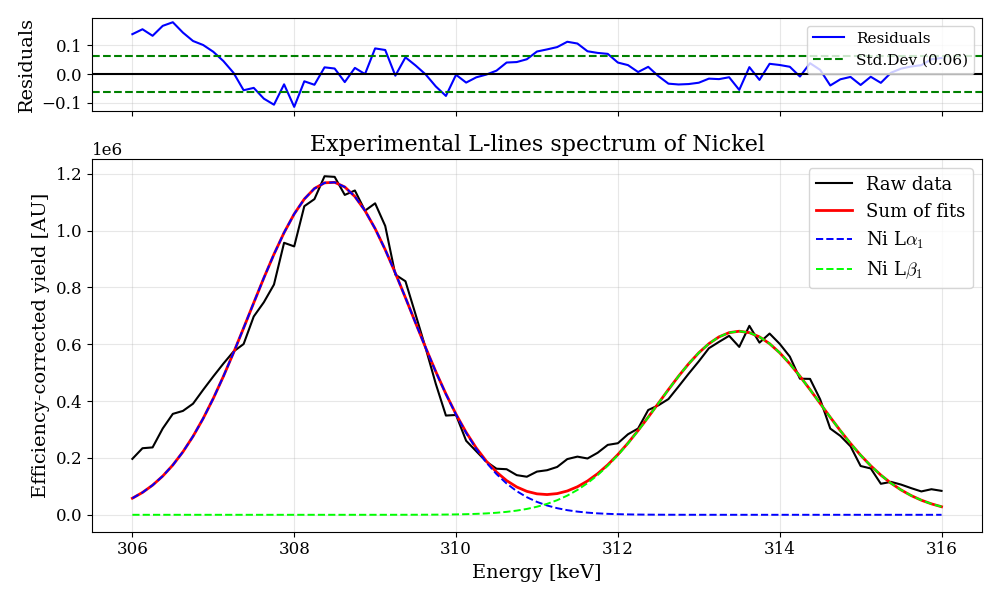}
        \vspace{4ex} 
    \end{subfigure}

    \begin{subfigure}[b]{1\linewidth}
        \centering
        \includegraphics[width=0.85\linewidth]{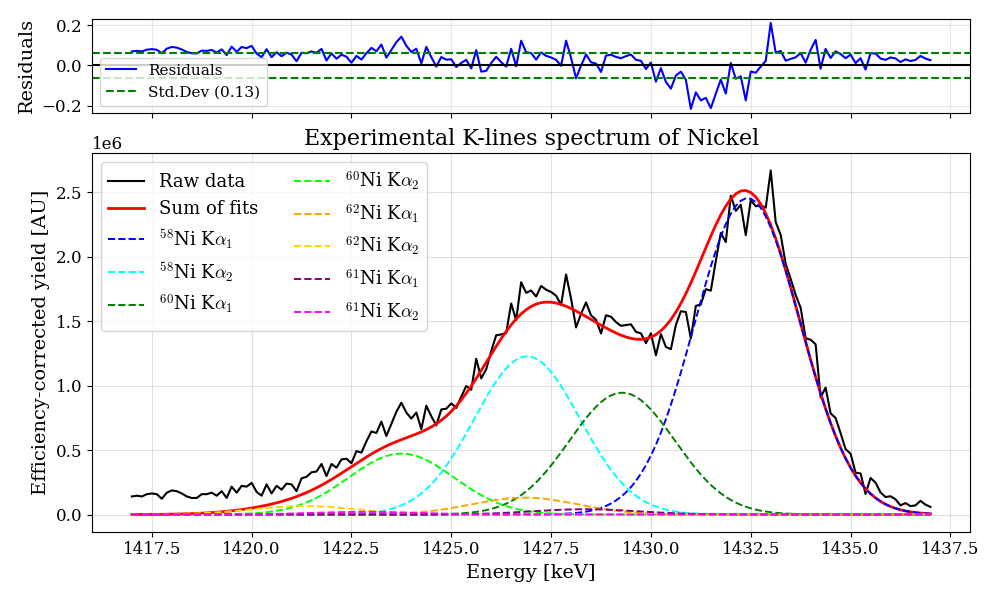}
        \vspace{4ex} 
    \end{subfigure}

    \caption{Measured nickel muonic x-ray emission spectrum as detected from the NMC layer of the battery at a momentum of 23.8~MeV/c. Isotopic shifts of $^{58}$Ni and $^{60}$Ni are clearly visible in the K-lines. Gaussian curves of each isotope K and L emission lines compose the function in red. Up: L lines. Down: K lines.}
    \label{fig:Ni_spectrum}
\end{figure}

The transition energies of nickel K-lines measured experimentally (Fig.~\ref{fig:Ni_spectrum}) agree closely with the predictions of MuDirac. A Gaussian decomposition performed on K and L lines groups (Tab.~\ref{tbl:Ni_fit}) shows consistent energy spacing across isotopes. The areas of these Gaussian curves will be used to estimate the ratio of intensities between K and L emission lines.

Nickel was selected as a test case because its relatively high atomic number leads to intense and well-separated muonic K transitions at energies above 1.4~MeV. At these energies, self-absorption within the sample is negligible and attenuation in the surrounding materials is minimal compared to the L lines. In addition, the natural isotopic composition of Nickel produces clearly resolved isotope-dependent shifts, providing a stringent test of the cascade and transport modeling.

\subsubsection{Nickel spectral fit and relative intensities}

The relative line intensities were extracted by modeling each transition with a Gaussian response function of fixed instrumental width $\sigma$, assuming detector resolution dominates over the intrinsic line width. For the K series, the amplitudes of the doublet components were constrained such that
\begin{align}
R_K = \frac{I(K\alpha_2)}{I(K\alpha_1)} = 0.5 ,
\end{align}
as obtained from PHITS and the Akylas cascade code. This constraint was imposed consistently for all Nickel isotopes, thereby reducing the number of free parameters and enforcing physical consistency of the fit. No analogous constraint was applied to the $\mathrm{L\alpha_1}$ and $\mathrm{L\beta_1}$ lines around 308 and 313~keV, as no universal fixed ratio was observed across the set of assessed elements.

For each Nickel isotope $i$, the fitted peak area $A_{i,\ell}$ satisfies
\begin{align}
A_{i,\ell} \propto N_i , Y_{i,\ell} , \varepsilon(E_\ell) ,
\end{align}
where $N_i$ is the number of stopped muons captured by isotope $i$, $Y_{i,\ell}$ the radiative yield of line $\ell$, and $\varepsilon(E_\ell)$ the full-energy peak detection efficiency at energy $E_\ell$. Because the K lines lie at sufficiently high energy, attenuation and self-absorption effects are negligible compared to the L lines, making the K series a robust normalization reference for testing cascade yields.

The global intensity ratio between K and L emission was therefore defined as

\begin{align}
\frac{K_\alpha}{L} =
\frac{\sum_i \left[A_{i,K\alpha_1} + A_{i,K\alpha_2}\right]}
{A_{L\alpha_1} + A_{L\beta_1}} .
\end{align}
It should be emphasized that the ratio defined above is not a purely atomic quantity. Because the detector efficiency, photon attenuation in the sample and surrounding materials, and the experimental geometry enter differently at 300~keV and 1.4~MeV, the observable $\mathrm{K_\alpha/L}$ is inherently geometry dependent. The comparison therefore probes not only the atomic cascade yields $\mathrm{Y_{i,\ell}}$ but also the consistency of photon transport and efficiency treatment in the simulation.
From the ensemble of Gaussian fits to the experimental spectrum retrieved from datasets of \cite{edouard}, compiled in Tab.~\ref{tbl:Ni_fit}, this ratio is found to be

\begin{align}
\left( \frac{K_\alpha}{L} \right)_{\mathrm{exp}} = 3.72 
\end{align}

The corresponding value obtained from PHITS is

\begin{align}
\left( \frac{K_\alpha}{L} \right)_{\mathrm{PHITS}} = 1.54 \pm 0.08 .
\end{align}

The quoted uncertainty reflects statistical propagation of the fitted peak-area uncertainties. The experimental ratio, however, is affected by additional systematic contributions that are not explicitly available in the published dataset. These include HPGe energy calibration and efficiency calibration uncertainties, geometrical alignment and solid-angle effects, dead-time corrections, as well as possible self-absorption and background-subtraction systematics. Consequently, the comparison should be interpreted primarily at the level of global consistency rather than as a precision benchmark.

Although PHITS underestimates the relative K-to-L emission strength, it reproduces the correct order of magnitude and preserves the expected hierarchy between high-energy and low-energy cascade transitions. Given the strong sensitivity of L-line intensities to atomic cascade modeling, low-energy attenuation, and detector efficiency corrections, this agreement provides a meaningful validation of the coupled cascade–transport treatment implemented in the code. Further refinement would require a fully efficiency-corrected experimental spectrum and a detailed treatment of low-energy photon transport in the experimental geometry.
\begin{table}[htb]
\centering
\caption{Parameters of Gaussian functions fitted to the Nickel spectra in Fig.~\ref{fig:Ni_spectrum}. The instrument resolution $\sigma$, in keV, was set to 1.32 (K-lines) and 1.10 (L-lines).}
\vspace{3mm}
\begin{tabular}{cccc}
\hline
Isotope & Line & Average (keV) & Area (AU) \\
\hline
$\mathrm{^{58}Ni}$ & $\mathrm{K\alpha_1}$ & 1432.42 & 8.13 \\
$\mathrm{^{58}Ni}$ & $\mathrm{K\alpha_2}$ & 1426.90 & 4.06 \\
$\mathrm{^{60}Ni}$ & $\mathrm{K\alpha_1}$ & 1429.28 & 3.13 \\
$\mathrm{^{60}Ni}$ & $\mathrm{K\alpha_2}$ & 1423.75 & 1.56 \\
$\mathrm{^{62}Ni}$ & $\mathrm{K\alpha_1}$ & 1426.84 & 0.43 \\
$\mathrm{^{62}Ni}$ & $\mathrm{K\alpha_2}$ & 1421.32 & 0.22 \\
$\mathrm{^{61}Ni}$ & $\mathrm{K\alpha_1}$ & 1428.36 & 0.14 \\
$\mathrm{^{61}Ni}$ & $\mathrm{K\alpha_2}$ & 1422.84 & 0.07 \\
Ni (all) & $\mathrm{L\alpha_1}$ & 308.34 & 3.07 \\
Ni (all) & $\mathrm{L\beta_1}$  & 313.30 & 1.69 \\
\label{tbl:Ni_fit}
\end{tabular}
\end{table}

\section{Conclusion}
A detailed benchmarking of GEANT4, PHITS, and SRIM has been conducted for the simulation of Muon-Induced X-ray Emission in a multilayer lithium-ion battery cell. GEANT4 and PHITS yield consistent predictions for stopping depths of muons, even in geometries with pronounced density contrasts, confirming their suitability for quantitative MIXE analysis. SRIM, while not originally intended for muon transport, remains a useful and rapid estimator of stopping ranges within individual layers, though its precision deteriorates when extended low-density materials such as air or other gases precede the target.

The modeling of muon capture and atomic cascades in PHITS reproduces relative line intensities with acceptable margins, but exhibits systematic energy discrepancies in K-line transitions of medium- to high-Z elements. This limitation currently prevents direct spectral comparison, yet the consistency of the predicted intensities underscores the reliability of the underlying cascade model. A practical remedy would be to couple PHITS to tabulated muonic-atom transition energies—either via lookup tables generated with MuDirac or by directly integrating the MuDirac routine into the PHITS atomic-cascade module—while retaining PHITS' legacy cascade logic to compute relative intensities. Such a hybrid approach would substantially improve predictive accuracy.

Overall, SRIM and PHITS constitute efficient, user-friendly tools for predicting muon ranges and beam-momentum requirements in layered materials, while PHITS additionally enables preliminary spectral simulation. These results support the ongoing development of a web-based simulation platform for future MIXE users and provide a foundation for improving muonic-cascade modeling in next-generation versions of PHITS.

\appendix
\renewcommand{\thetable}{A.\arabic{table}} 
\setcounter{table}{0}                      

\section*{Appendix A: Supplementary Data}
\addcontentsline{toc}{section}{Appendix A: Supplementary Data}

\begin{table}[htb]
	\centering 
	\caption{Key physics cards used in the set of PHITS simulations of the lithium-battery case.}
	\vspace{3mm} 
	\begin{tabular}{ c c c }
		Card & Parameter & Description \\
		\hline
		\hline
		emumin & 1.0e-3 & minimum muon energy for nuclear reactions, in MeV \\
		
		nspred & 1 & option for Coulomb diffusion (angle straggling)\\
		
		nedisp & 1 & Energy straggling option for charges particles and nuclei \\
		
		delt0 & 3e-4 & max. step size if nspred =1 \\
		
		imubrm & 0 & muon-induced bremsstrahlung \\
		
		imuppd & 1 & muon-induced pair-production \\
		
		imuint & 1 & muon induced nuclear reaction with virtual photon theory \\
		
		imucap & 1 & negative muon capture reaction and muonic X-ray emission \\
		
		igamma & 0 & gamma photons generation and transport \\
		
		\label{tbl:physics}
	\end{tabular}
\end{table}

\begin{table}[htb]
    \centering
    \caption{%
        Relative difference in energy between PHITS and MuDirac for various isotopes. 
        Values are expressed as percentage difference from MuDirac reference energies. 
        PHITS values for heavy elements such as gold or lead could not be computed.}
    \vspace{3mm}
    \begin{tabular}{lcc}
        \hline
        \textbf{Isotope} & \textbf{$\Delta K_\alpha$ line (\%)} & \textbf{$\Delta L_\alpha$ line (\%)} \\
        \hline\hline
        $\mathrm{^{7}Li}$   & -0.18   & -0.16 \\
        $\mathrm{^{12}C}$   &  0.04   & -0.16 \\
        $\mathrm{^{27}Al}$  &  2.64   & -0.33 \\
        $\mathrm{^{28}Si}$  &  3.23   & -0.35 \\
        $\mathrm{^{39}K}$   &  7.15   & -0.43 \\
        $\mathrm{^{55}Mn}$  & 13.40   & -0.47 \\
        $\mathrm{^{58}Ni}$  & 16.79   & -0.47 \\
        $\mathrm{^{59}Co}$  & 15.90   & -0.47 \\
        $\mathrm{^{63}Cu}$  & 18.65   & -0.46 \\
        $\mathrm{^{74}Ge}$  & 23.72   & -0.42 \\
        $\mathrm{^{79}Br}$  & 28.37   & -0.36 \\
        $\mathrm{^{89}Y}$   & 34.85   & -0.24 \\
        $\mathrm{^{120}Sn}$ & 59.45   &  0.55 \\
        $\mathrm{^{133}Cs}$ & 72.42   &  1.20 \\
        $\mathrm{^{159}Tb}$ &102.24   &  3.05 \\
        $\mathrm{^{208}Pb}$ &   ---   &  8.62 \\
        \hline
    \end{tabular}
    \label{tbl:appendix_shift}
\end{table}
	\begin{figure}[ht] 
		\begin{subfigure}[b]{0.5\linewidth}
			\centering
			\includegraphics[width=0.95\linewidth]{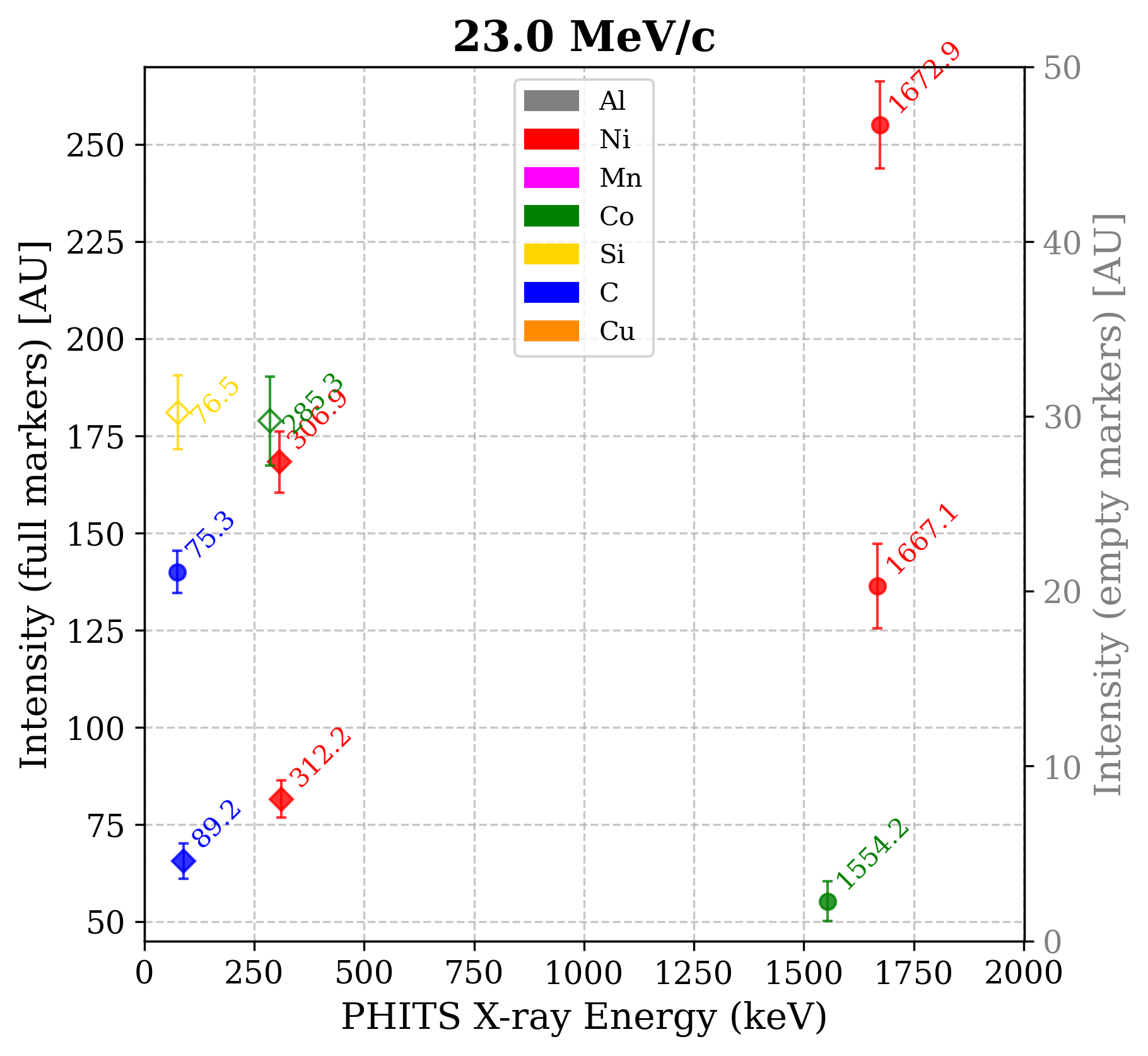} 
			\vspace{4ex}
		\end{subfigure}
		\begin{subfigure}[b]{0.5\linewidth}
			\centering
			\includegraphics[width=0.95\linewidth]{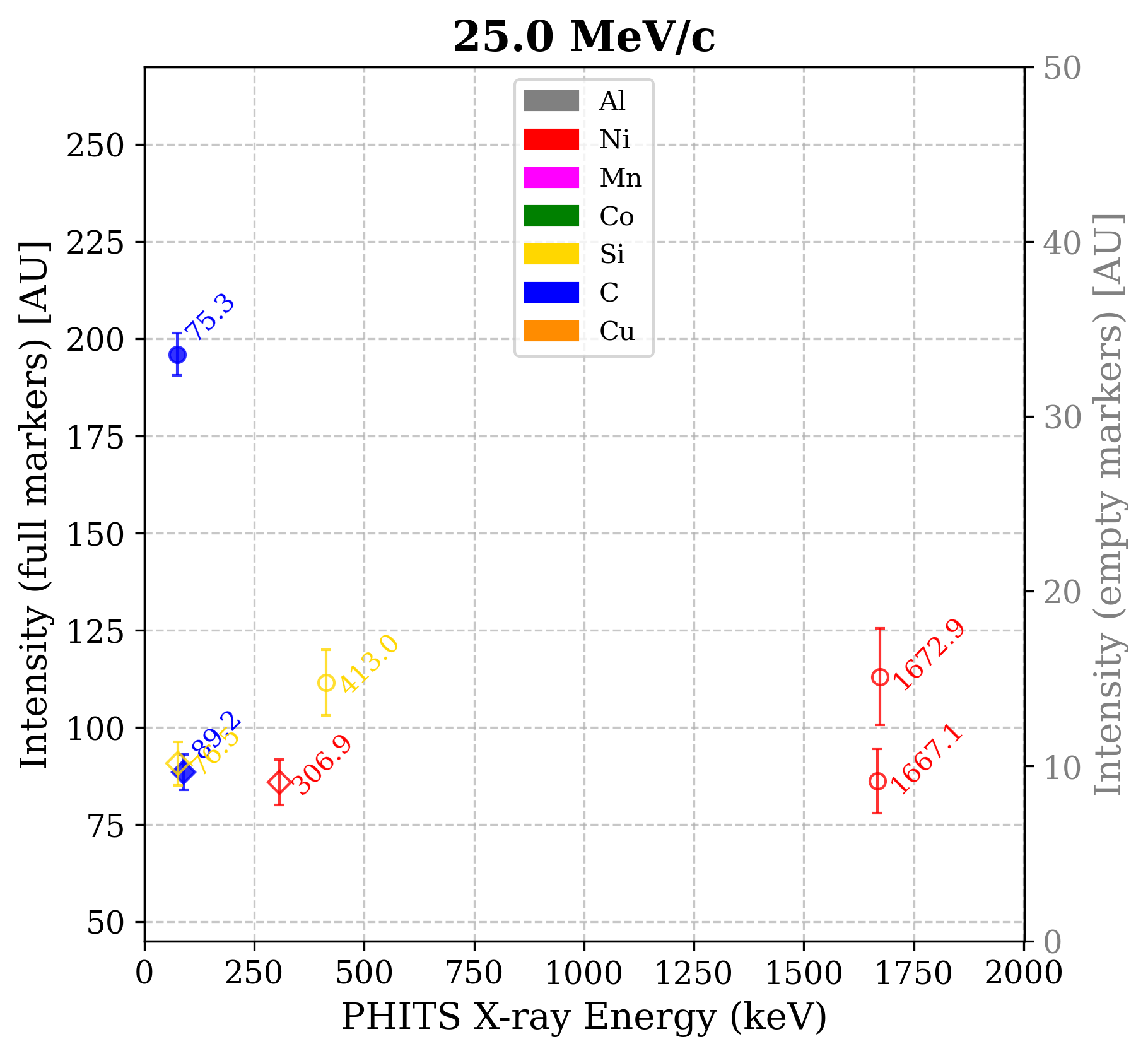} 
			\vspace{4ex}
		\end{subfigure} 
		\begin{subfigure}[b]{0.5\linewidth}
			\centering
			\includegraphics[width=0.95\linewidth]{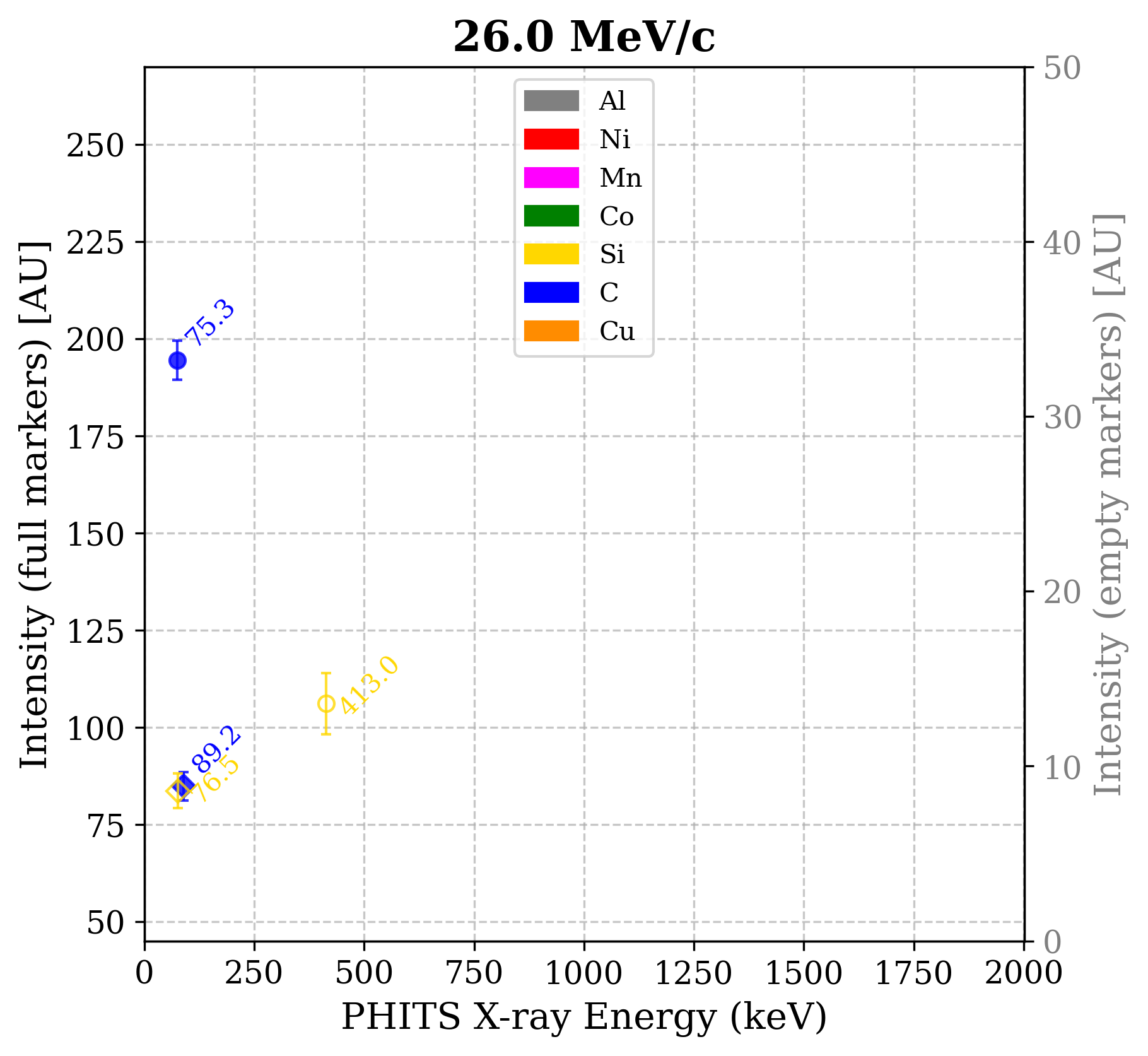} 
			\vspace{4ex}
		\end{subfigure} 
		\begin{subfigure}[b]{0.5\linewidth}
			\centering
			\includegraphics[width=0.95\linewidth]{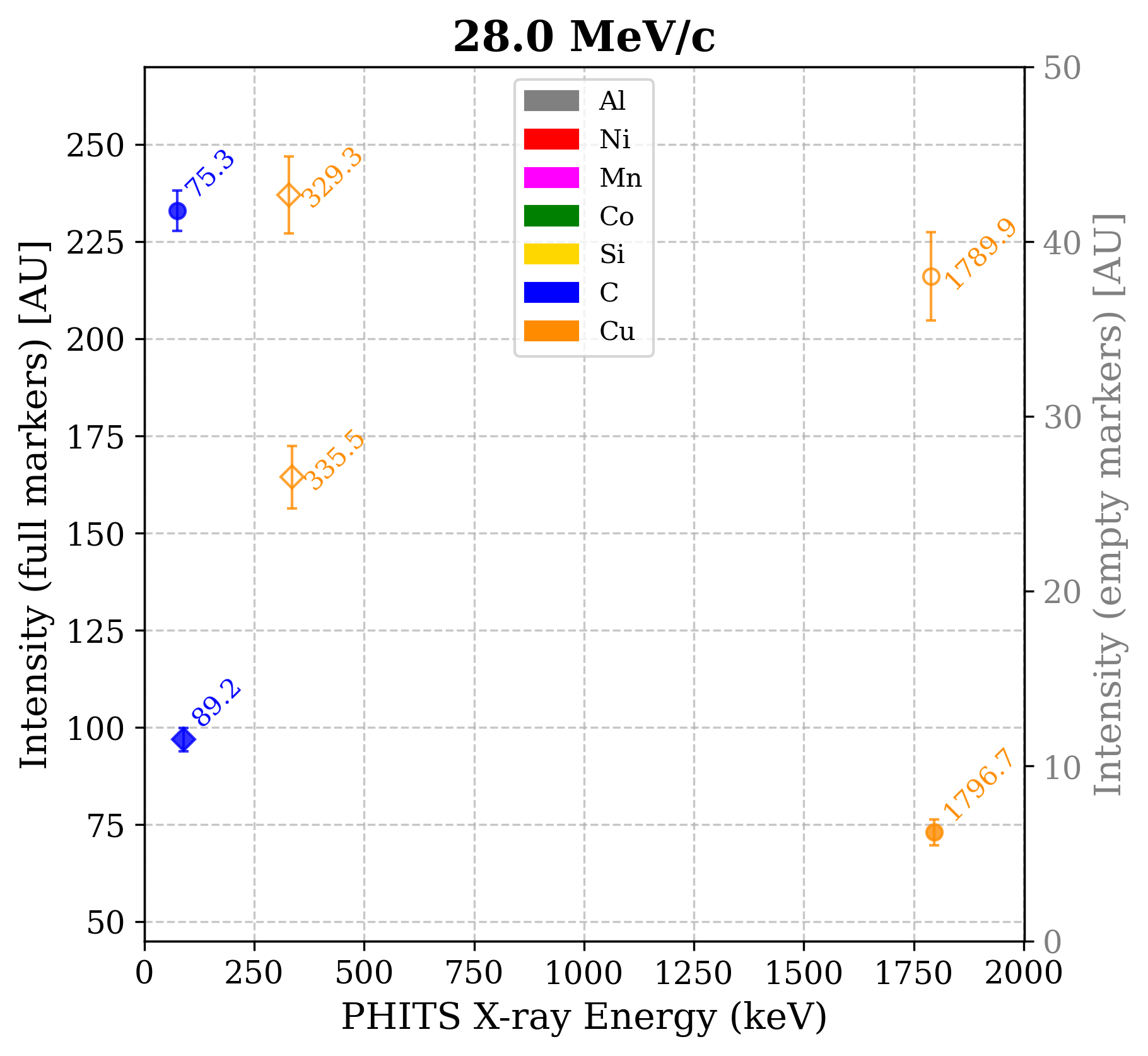} 
			\vspace{4ex}
		\end{subfigure} 	
		\caption{MIXE signal as estimated by PHITS for various momenta in the test battery. Each layer constitutive element should appear at a given momenta/stopping depth, such as estimated on Fig.~\ref{fig:profiles_all}. Error bars are 1$\sigma$ Monte-Carlo statistical uncertainties from PHITS. A nominal 0.5 keV resolution ($\approx$ half a tally bin) is assumed to merge unresolved transitions and to define the energy integration window across adjacent bins.}
		\label{fig:mixe_all}
	\end{figure}
\clearpage
\bibliography{bibPHITS.bib}
	
\end{document}